\documentclass[prd, aps, superscriptaddress, preprintnumbers, twocolumn, nofootinbib, floatfix]{revtex4-2}
\pdfoutput=1

\usepackage{array}
\usepackage{amsmath}
\usepackage{amsfonts}
\usepackage{amssymb}
\usepackage{bm}
\usepackage{bbm}
\usepackage{cases}
\usepackage{cancel}
\usepackage{dcolumn}
\usepackage{graphicx}
\usepackage{hyperref}
\usepackage[latin1]{inputenc}
\usepackage{multirow}
\usepackage{subfigure}
\usepackage{setspace}
\usepackage{ulem}
\usepackage{xcolor}

\newcommand{\nn}{\nonumber \\}

\newcommand{\cR}{{\cal R}}

\newcommand{\gphi}{g_{\phi\gamma}}
\newcommand{\F}{F_{\mu\nu}}
\newcommand{\gt}{{\tilde{g}_{10}}}
\newcommand{\gsim}{\mathrel{\hbox{\rlap{\lower.55ex \hbox {$\sim$}}
                   \kern-.3em \raise.4ex \hbox{$>$}}}}
\newcommand{\lsim}{\mathrel{\hbox{\rlap{\lower.55ex \hbox {$\sim$}}
                   \kern-.3em \raise.4ex \hbox{$<$}}}}

\newcommand\be{\begin{equation}}
\newcommand\bt{\begin{table}}
\newcommand\bts{\begin{table*}}
\newcommand\bfig{\begin{figure}}
\newcommand\bfs{\begin{figure*}}
\newcommand\bi{\begin{itemize}}
\newcommand\ee{\end{equation}}
\newcommand\et{\end{table}}
\newcommand\ets{\end{table*}}
\newcommand\efig{\end{figure}}
\newcommand\efs{\end{figure*}}
\newcommand\ei{\end{itemize}}

\newcommand{\bal}{\begin{align}}
\newcommand{\eal}{\end{align}}

\newcommand\tbf{\textbf}
\newcommand\tit{\textit}

\definecolor{llgray}{gray}{0.88}

\begin{document}

\title{Constraints on Axion-photon coupling from the Global 21-cm Signal}

\author{Hao Jiao}
\email{jiaohao@ibs.re.kr}
\affiliation{Department of Physics, McGill University, Montr\'{e}al, QC, H3A 2T8, Canada}
\affiliation{Cosmology, Gravity and Astroparticle Physics Group, Center for Theoretical Physics of the
Universe, Institute for Basic Science (IBS), Daejeon, 34126, Korea} 

\date{\today}


\begin{abstract}

A radiation field can be excited via parametric resonance when an oscillating axion field couples to the electromagnetic sector through a Chern-Simons interaction. As demonstrated in previous works, this mechanism can generate primordial magnetic fields shortly after recombination and provide sufficient ultraviolet radiation for the formation of direct collapse black holes (DCBHs). In this study, I analyze constraints on the parametric resonance scenario from global 21cm observations. I find that there exist viable regions in the parameter space that satisfy both observational limits and the physical requirements of the magnetic field and DCBH formation scenarios.

\end{abstract}


\maketitle

\section{Introducton}
\label{sec1}

The pseudo-scalar axion field is a well-motivated candidate for dark matter, originally proposed to solve the strong CP problem in QCD \cite{axion-1,axion-2} and can also play a role in explaining various unresolved aspects of cosmology \cite{axion-3,axion-4}. In this work, I am interested in ultra-light axion-like particles \cite{FDM-1,FDM-2,FDM-3}, which could behave as a coherent classical field, oscillating around the minimum of its potential after decoupling from the Hubble expansion \cite{RB-misalignment,misalignment-2,misalignment-3,misalignment-4,misalignment-5}. In particular, it oscillates with a frequency set by its mass, i.e., $\phi=\phi_0 \sin(mt)$, where $m$ is the axion mass and $\phi_0$ is the initial amplitude determined by the energy density of axion dark matter.

When coupled to the electromagnetic U(1) gauge field via a Chern-Simons term, the oscillating axion background will induce broad parametric resonance in the gauge field. This coupling breaks the scale invariance and enables energy transfer from the axion to the electromagnetic sector \cite{violate-1,violate-2,violate-3,JF1,JF2,Joyce,RB-magnetic}. As a result, modes of the gauge field with wavenumber smaller than a critical value can undergo an exponential amplification. 
The parametric resonance of the tachyonic type\footnote{For half of the oscillation period of the axion field, the effective mass term in the equation of motion for the gauge field is negative, leading to a ``tachyonic'' instability.} occurring in the very early universe has been studied previously \cite{Evan1,Evan2,Rudnei}. 
Recently, the broad resonance at later times has attracted attention due to its potential to explain various cosmological phenomena, such as the origin of large-scale primordial magnetic fields \cite{RB-magnetic} and the formation of high-redshift supermassive black hole (SMBH) seeds \cite{JH-DCBH}.

The magnetic fields on intergalactic scales have been detected by various observations \cite{PMF-1,PMF-2}, which are unlikely to originate from known astrophysical processes. They are therefore often interpreted as primordial magnetic fields. The generation mechanism of such fields remains under debate.  
Popular scenarios involve scale-invariance-breaking processes during inflation, typically through a Chern-Simons coupling between the inflaton and a $U(1)$ gauge field \cite{PMF-gen-1,PMF-gen-2,PMF-gen-3,PMF-gen-4}. However, these models face the issue of a significant decay in the magnetic field amplitude due to the vast cosmic expansion from inflation to the present day. 
In \cite{RB-magnetic}, we propose a scenario in which strong magnetic fields on cosmological scales can be generated via a ``tachyonic'' type parametric resonance shortly after recombination. This occurs when electromagnetism is coupled to an ultralight pseudo-scalar axion field (fuzzy dark matter), which coherently oscillates on cosmological scales, through the standard $\phi F\wedge F$ interaction. This mechanism is capable of explaining the observed magnetic fields with reasonable model parameters.

Observations of high-redshift quasars indicate the presence of supermassive black holes (SMBHs) with masses exceeding $10^9 M_\odot$ at redshifts $z > 6$ \cite{AGN1,AGN2,AGN3,AGN4,AGN5}. The origin of these SMBH seeds remains a mystery. An important formation channel is the direct collapse of atomic cooling gas clouds, which can generate ``heavy'' seeds with typical masses of $\sim 10^5 M_\odot$ \cite{DCBH-1,DCBH-2,DCBH-3,DCBH-4,DCBH-5,DCBH-6,DCBH-7,DCBH-8,DCBH-9}. A key requirement for this DCBH formation scenario is the presence of intense Lyman-Werner (LW) radiation, which suppresses $H_2$ formation and thus prevents the fragmentation of gas clouds. In the context of standard astrophysical models, such radiation can only be generated by stars.
In \cite{JH-DCBH}, we demonstrate that resonantly excited radiation from an oscillating axion dark matter field can provide a sufficient LW background to satisfy the DCBH formation conditions, thereby enabling the formation of massive black hole seeds before the onset of star formation in halos.

The parametric resonance mechanism leads to exponential growth of the electromagnetic field amplitude until the backreaction effect quenches the amplification, which is assumed to occur when a fraction $f$ of the axion energy is converted into radiation. To ensure consistency with observations, we aim to derive constraints on this quenching energy fraction $f$.

The global 21cm signal, which probes the hyperfine transition of neutral hydrogen, is highly sensitive to the background radiation field \cite{21cm-1,21cm-2}. Extra resonant radiation will increase the background radiation temperature, and thereby deepen the absorption feature in 21cm lines. Current measurements, such as those reported by the EDGES collaboration \cite{EDGES}, can thus be used to constrain such excess radiation. In this work, I discuss the constraints on the parametric resonance scenario from global 21cm observations.

In the following section, I review the parametric resonance of the electromagnetic gauge field coupled to an oscillating axion field, along with the mechanisms that can broaden the spectrum of the excited radiation. Section III presents constraints on the radiation field from global 21cm observations, considering both the global resonance shortly after recombination, when the axion field oscillates coherently on cosmological scales, and the resonance occurring within halos. Finally, Section IV presents the main conclusions with a discussion of these constraints and the relevant parametric resonance scenarios.

Here I work in natural units with $c = k_B = \hbar = 1$ and express all quantities in units of $GeV$ unless otherwise noted. The photon energy (and corresponding wavenumber) is given in units of $eV$. I consider a homogeneous and isotropic expanding universe described by the Friedmann-Lemaitre-Robertson-Walker metric, with the scale factor normalized such that $a(t_0)=1$, where $t_0$ denotes the present time. Accordingly, the scale factor at redshift $z$ is $a(t)=1/(z+1)$.

\section{Resonant production of the radiation field}
\label{sec2}

The Lagrangian of the gauge field $A_\mu$ that couples to the axion field $\phi$ by a Chern-Simons term is 
\be
{\cal{L}} \, = \, -\frac12 (\partial\phi)^2 + V(\phi) -\frac14 F_{\mu\nu}F^{\mu\nu} + g_{\phi\gamma}\phi F_{\mu\nu}\tilde{F}^{\mu\nu}, \label{eq-Lagrangian}
\ee
where $\F \equiv \partial_\mu A_\nu - \partial_\nu A_\mu$ and $\tilde{F}_{\mu\nu}$ is the dual of $\F$.  $\gphi$ is the coupling constant of units $GeV^{-1}$, and its current upper bound is $\gphi \gsim 10^{-10}GeV^{-1}$ \cite{axion-3, axionphoton-1}. Hence, I define the dimensionless coupling constant $\gt \equiv \frac{\gphi}{10^{-10}GeV^{-1}}$ for simplification in the following calculations.

Here we assume that the field $\phi$ corresponds to axion-like ultralight dark matter particles with mass $m$,\footnote{The axion mass should be in the range of $10^{-20}\text{eV} < m < 10 eV$  \cite{AxionMass1, AxionMass2}}. The potential $V(\phi)$ is approximately quadratic around the minimum $\phi\rightarrow0$
\be
V(\phi) \, \simeq \, \frac12 m^2\phi^2 \, ,
\ee 
resulting in an oscillatory solution
\be
\phi=\phi_0 \sin(mt) \label{eq-phi-sol}
\ee 
when Hubble friction and spatial gradients are negligible compared to the time variation of the field. The amplitude $\phi_0$ is determined by the energy density of dark matter with the relation
\be
\rho_{axion} \sim \frac12 m^2\phi_0^2.
\ee

This solution will lead to a ``tachyonic'' resonance of the electromagnetic field in an instability region $k<k_c$, yielding significant growth of radiation fields. In the next two subsections, I will discuss the parametric resonance of the global radiation field and the radiation field in halos separately.

\subsection{Parameteric resonance of the global radiation}
\label{sec2-1}

In the early universe, the axion field begins to oscillate coherently after decoupling from the Hubble flow, leading to the generation of a global electromagnetic field shortly after recombination. This mechanism provides a potential explanation for the origin of primordial magnetic fields on scales of 1 Mpc \cite{RB-magnetic}, and I briefly introduce it in this subsection.

In Fourier space, the comoving helicity modes of the electromagnetic gauge field, $A_k^\pm$, satisfies the equation of motion (EOM) derived from the Lagrangian in eq.~\eqref{eq-Lagrangian} \cite{RB-magnetic, magnetic-1}.
\be
\Big[\partial_\eta^2 + {k^{(c)}}^2 \pm 4g_{\phi\gamma} {k^{(c)}} a(\eta) \dot\phi \Big] A_k^\pm \, = \, 0 \, , \label{eq-EOM-1}
\ee
where we use the conformal time $\eta$ to absorb the Hubble damping term and ${k^{(c)}}$ is the comoving wavenumber. For the oscillating axion field solution given in eq.~\eqref{eq-phi-sol}, we can approximate $\dot\phi\simeq m\phi_0\cos(mt)$ by neglecting the time variation of $\phi_0$ over an oscillation period. Therefore, in the broad instability region 
\begin{align}
{k^{(c)}} < k^{(c)}_c \equiv 4 g_{\phi\gamma}a m \phi_0,
\end{align}
the gauge field modes will undergo an exponential growth in their amplitude with the Floquet exponent\footnote{The Floquet exponent characterizes the exponential growth of the amplitude $|A_k^\pm| \propto \exp(\mu_k \Delta t)$ during the ``tachyonic'' phase within each oscillation period, i.e. when the effective mass term $m_{eff}^2 = {k^{(c)}}^2 \pm 4g_{\phi\gamma} {k^{(c)}} a \dot\phi \sim \pm 4g_{\phi\gamma} {k^{(c)}} a m \phi_0 \cos(mt)$ is negative, neglecting the $k^2$ contribution.}
\be
\mu_k \sim \sqrt{2 g_{\phi\gamma}a m \phi_0 k^{(c)}} \, , 
\ee
which is significantly larger than the Hubble parameter for the coupling constant $g_{\phi\gamma}$ in the range of interest \cite{RB-magnetic}. Thus, the global radiation will be significantly enhanced in a short period of time, until the backreaction becomes important. We assume that the quenching occurs when a fraction $f$ of axion energy converts into radiation, and this fraction is an important parameter that needs to be constrained by observations.

As the Floquet exponent increases with the increase of $k^{(c)}$ for $k^{(c)} < k^{(c)}_c$, the amplification of the electromagnetic field is most efficient near the cutoff scale $k^{(c)}_c$, resulting in a pronounced peak in the radiation spectrum around this wavenumber.

Note that the EOM in eq.~\eqref{eq-EOM-1} is only valid after recombination, when the radiation field is not affected by the baryon-photon plasma. Therefore, the parametric resonance is expected to occur shortly after recombination. At this time, the amplitude of the axion field is roughly
\be
m\phi(t_{rec}) \sim T_{rec}^2,
\ee
and the corresponding critical wavenumber is
\be
k^{(c)}_c = 3\times10^{-23} eV \cdot \tilde{g}_{10}.
\ee
Note that we use comoving wavenumber here, so the corresponding photon energy is
\be
k_{c} = k^{(c)}_c/a = 4 g_{\phi\gamma} m \phi_0 =  3\times10^{-20}eV \cdot \tilde{g}_{10}.
\ee

\subsection{Parameteric resonance in halos}
\label{sec2-2}

The parametric resonance can also happen in halos, where the bound axion field oscillates coherently after virialization. The strong radiation field sourced by this oscillation could play an important role in the formation scenario of DCBHs, which are candidates of SMBH seeds \cite{JH-DCBH}. The resonance mechanism discussed here is similar to the one described above. However, note that the two processes are independent, so we should analyze the constraints separately. 

In halos that consist of an oscillating dark matter field, the EOM of $A_k$ is given by
\be
\Big( \partial_t^2 + k^2 \pm 4g_{\phi\gamma} k m\phi_0 \cos(mt) \Big)A_k \, = \, 0 \, . \label{eq-A-EOM}
\ee
In the bounded system, we use physical time $t$ and physical wavenumber. Thus, strong radiation will be produced in the halo with photon energy 
\be
k \sim k_c \equiv 4\gphi m\phi_0,
\ee
which is determined by the halo dark matter energy density
\be
\rho_{axion} \sim \Delta_{vir} \rho_{bg}(z) \sim (m\phi_0)^2 \, .
\ee

In this scenario, we still assume that a fraction $f$ of dark matter energy converts to radiation during the resonance. But since the backreaction mechanisms in the two parametric resonance scenarios might differ, the constraints on the quenching energy fraction $f$ are independent.

Additionally, in contrast to the previous global radiation, only a fraction $\epsilon$ of radiation energy generated in halos could escape and influence the global 21cm signal. 
Note that, in principle, $\epsilon$ varies for different photon frequencies, but as I am only interested in radio photons that could contribute to 21cm signals after redshifting, I neglect the frequency dependence on the escape fraction.

\subsection{Energy spectrum of the radiation}
\label{sec2-3}

Parametric resonance would lead to strongly excited radiation sharply peaked around $k_c$, which corresponds to very low-frequency photons. In what follows, I introduce two mechanisms that can transfer the narrow excited spectrum to a broader one, i.e., energy cascade and thermalization.

With turbulence, the energy of low-frequency photons can be transferred to high-energy photons through an energy cascade, leading to a power-law energy spectrum
\be
\frac{d\rho_{ec}}{dk} \, = \, \frac{f\rho_{axion}}{k_c}\bigg(\frac{k_c}{k}\bigg)^n \, , \label{eq-cascade}
\ee
where $n$ is the scaling index, and we treat it as a free parameter that should be constrained.\footnote{The value of $n$ depends on the detail of the turbulent system. For Kolmogorov scaling we have $n = 5/3$ \cite{Kolmogorov}, while for Batchelor scaling $n = 1$ \cite{Batchelor}. In general, $n$ should be larger than 1 to ensure the convergence of the total energy.}

On the other hand, if the excited radiation is efficiently thermalized in gas clouds, it will follow a black-body spectrum:
\be
\frac{d\rho_{bb}}{dk} = \frac{k^3}{4\pi^3}\frac{1}{e^{k/T_{bb}}-1}, \label{eq-thermal}
\ee 
where $T_{bb} = \big(f\rho_{axion}\big)^{1/4}$ is the temperature of the radiation in thermal equilibrium.

\section{Constraints from global 21cm signals}
\label{sec3}

The excited radiation driven by the oscillating axion field could lead to a radio excess in the intergalactic medium (IGM) and amplify the absorption in the global 21cm signal. This feature is characterized by differential brightness temperature, which quantifies the difference between the observed brightness temperature of the 21cm line and the CMB background radiation temperature. The EDGES collaboration reported a strong absorption feature that is roughly twice as deep as the prediction of the $\Lambda$CDM model at redshift $z \sim 17$ \cite{EDGES}.\footnote{However, this result remains under debate due to the no detection of this feature using SARAS 3 radiometer \cite{EDGES-problem}.}  

A constraint on the parametric resonance mechanism can be determined by requiring the extra radiation not to conflict with the global 21cm observations before reionization. 
In a mostly neutral IGM, the differential brightness temperature is approximately proportional to 
\begin{align}
\delta T_b \, \propto \, \bigg( 1-\frac{T_\gamma}{T_s} \bigg),
\end{align}
where $T_s$ is the spin temperature of the hydrogen gas, and $T_\gamma$ is the equivalent temperature of the photons at 21cm frequency, including both CMB photons and the extra resonant radiation
\begin{align}
T_{\gamma} \, &= \, T_{CMB} + T_{rad}(k_{21})\\
&= \, T_{CMB} ( 1 + \cR ), 
\end{align}
where $\cR$ is the ratio between the spectral energy density (per unit wavenumber) of the extra radiation to that of CMB at $k=k_{21}\equiv \frac{2\pi}{21 cm}=6\times10^{-6}eV$
\be
\cR \, \equiv \, \frac{d\rho_{rad}}{dk}(k_{21}) \bigg/ \frac{d\rho_{CMB}}{dk}(k_{21}).
\ee

For the CMB spectrum, the spectral energy density at 21cm frequency is
\begin{align}
\frac{d\rho_{CMB}}{dk}(k_{21}) \, &= \, \frac{k_{21}^3}{4\pi^3}\frac{1}{e^{k_{21}/T_{CMB}}-1}\\
&\simeq \, \frac{k_{21}^2}{4\pi^3} T_{CMB}(z)\\
&= \, 6.6\times10^{-44}GeV^3\cdot (z+1).
\end{align}

The excess radiation must be sufficiently small so as not to produce an absorption feature exceeding observational bounds, giving rise to the constraint
\be
\cR < 1.
\ee
In the following, I will compute $\cR$ for the two scenarios described in section \ref{sec2}.

\subsection{Constraint on the global resonance}
\label{sec3-1}

For the global radiation excited by parametric resonance, the initially narrow spectrum is primarily broadened through an energy cascade, because the thermalization of radiation only occurs in dense regions where interactions between photons and baryons are significant. Hence, I consider the power-law spectrum in eq.~\eqref{eq-cascade}.

The excited radiation was generated shortly after recombination, so the wavenumber $k$ of a photon detected at a redshift $z$ is related to the initial wavenumber $k'$ by
\be
k = k'\bigg(\frac{z+1}{z_{rec}+1}\bigg),
\ee
and the spectral energy density of the resonant radiation at the observed redshift $z$ is given by
\begin{align}
\frac{d\rho_{rad}}{dk} \, &= \, \frac{d\rho_{ec}}{dk'}\bigg|_{rec}\big(k'=k(\tfrac{z_{rec}+1}{z+1})\big)\bigg(\frac{z+1}{z_{rec}+1}\bigg)^3 \nn
&= \, \frac{f m\phi_0}{4\gphi} \bigg(\frac{4\gphi m\phi_0}{k}\frac{z+1}{z_{rec}+1}\bigg)^{n} \bigg(\frac{z+1}{z_{rec}+1}\bigg)^3,
\end{align}
where $(\frac{z+1}{z_{rec}+1})^3$ is the Jacobian due to the redshifting of the energy density, where $\rho_{rad}$ contributes a factor $(\frac{z+1}{z_{rec}+1})^4$ and the $dk$ in the denominator contributes a factor $(\frac{z+1}{z_{rec}+1})^{-1}$.

\begin{figure}
  	\includegraphics[width=7cm]{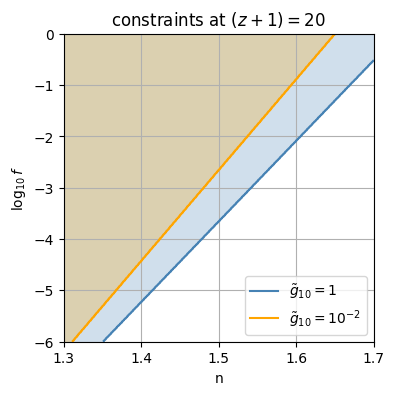}
	\caption{Constraints on the $n-f$ plane from global 21cm signal at redshift $(z+1)=20$ The shaded regions represent excluded parameter space for coupling constant $\gt=1,$ (blue) and $10^{-4}$ (orange).}
	\label{fig1}
\end{figure}

Thus, the ratio $\cR$ is
\begin{align}
\cR \, &= \, 4\pi^3 f \, (m\phi_0)^{n+1} (4\gphi)^{n-1} k_{21}^{-n-2} T_{rec}^{-1} \bigg(\frac{z+1}{z_{rec}+1}\bigg)^{n+2} \nn
&\approx \, 10^{-17.6 n + 25.2} \cdot f \, (4\tilde{g}_{10})^{n-1} (z+1)^{n+2}.
\end{align}
This ratio increases with redshift, and since the observed global 21cm absorption feature mainly arises during the cosmic dawn, I choose $(z+1)=20$ as a representative redshift to derive constraints on this scenario. The results are shown in Fig.~\ref{fig1}. 

In this figure, shaded areas correspond to regions in parameter space that are observationally excluded. The constraint on the radiation energy fraction becomes weaker with the increase of the index $n$ and the decrease of the coupling constant. The former is because 21cm photons have energy higher than $k_c$, and a larger $n$ corresponds to a steeper decline of the power-law spectrum at high energies, leading to weaker 21cm signatures. On the other hand, in contrast to our intuition, the overall strength of the excited radiation is independent of the value of the coupling constant $\gphi$. However, $k_c$ is proportional to $\gphi$, so increasing the coupling shifts the spectral peak to higher energies. As a result, the radiation at $k_{21}'$ will be stronger for larger $\gphi$.

\begin{figure}
  	\includegraphics[width=7cm]{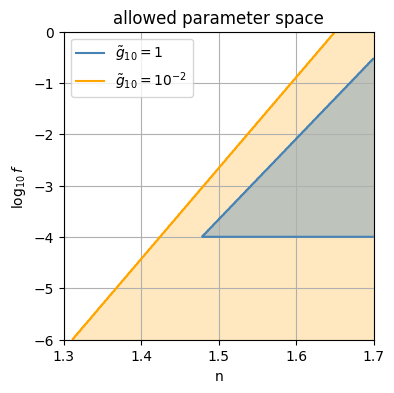}
	\caption{The shaded regions indicate the allowed parameter space where the generation of resonant magnetic field is consistent with global 21cm constraints. The color scheme is the same as in Fig. \ref{fig1}.}
	\label{fig2}
\end{figure}

This scenario could also generate a magnetic field with strength\footnote{{The helical magnetic field generated by parametric resonance could undergo an \tit{inverse} energy cascade, transferring energy from the critical mode to modes of larger scales $(k<k_c)$, 
\be
B(k) \simeq B(k_c) \bigg(\frac{k}{k_c}\bigg)^{3/2},
\ee
where the index $3/2$ characterizes the scaling of the inverse energy cascade and is not the same as the spectral index $n$ in this work.}}
\be
B \sim \gt^{-3/2} f^{1/2} 10^{15} \text{Gauss}
\ee
on length scale $\sim 1\text{Mpc}$ \cite{RB-magnetic}. Thus, to generate the primordial magnetic field of $B\sim 10^{-17}\text{Gauss}$, the radiation energy fraction $f$ must satisfy the condition
\be
f > 10^{-4} \gt^3 .
\ee 
The regions of parameter space that satisfy both the observational constraints as well as the above condition are shown in Fig.~\ref{fig2}. We find that the global 21cm constraint does not conflict with the requirement for magnetic field generation. Since the magnetic field scales as $B\propto\gphi^{-3/2}$, the viable region becomes significantly larger for smaller values of the coupling constant.

Before turning to parametric resonance in halos, I simply estimate the constraint under the assumption of a thermalized spectrum. But note that the excited global radiation is unlikely to be fully thermalized on cosmological scales after recombination. For the values of $f$ of interest,\footnote{This statement holds for $f> 10^{-40}$.} the radiation temperature is significantly larger than $k_{21}$, so the blackbody spectrum can be approximated similarly to the CMB spectrum
\begin{align}
\frac{d\rho_{rad}}{dk}(k_{21}) \, &\simeq \, \frac{k_{21}^2}{4\pi^3} T_{rad}(z_{rec})\cdot\bigg(\frac{z+1}{z_{rec}+1}\bigg).
\end{align}
Therefore, the ratio $\cR$ is equal to the ratio of the two radiation temperatures, which is roughly $\rho_{rad/CMB}^{1/4}$, i.e.,
\be
\cR \, = \, \frac{\rho_{rad}^{1/4}}{\rho_{CMB}^{1/4}}  = \bigg(\frac{f\rho_{axion}}{\rho_{CMB}}\bigg)^{1/4}
\ee 
At recombination, the energy density of CMB photons is one-third of the total matter density, and thus, $\cR<1$ leads to the constraint
\be
f < \frac13.
\ee
This constraint is relatively weak, as the backreaction effect is expected to already become significant when $f \ll 1$.

\subsection{Constraint on the resonance in halos}
\label{sec3-2}

For this scenario, I assume that the parametric resonance happens once per halo (shortly after virialization), so the injection rate of radiation energy into the background at a given time $t'$ is
\begin{align}
\frac{d\rho_{rad}}{dk'dt'} \, = \, \int dM\,\frac{dn}{dMdt'} \cdot \frac{dE}{dk'}(M,t')
\end{align}
where $M$ is halo mass and $\frac{dn}{dM\,dt'}$ is the corresponding halo formation rate at $t'$. Here, I adopt the Press-Schecter formalism to compute the halo mass function $\frac{dn}{dM}$, and $\frac{dn}{dM dt}$ is its time derivative. 
$\frac{dE}{dk'}$ is the escaped radiation energy (per unit wavenumber) from a single halo with mass $M$
\be
\frac{dE}{dk'}(M,t') \, = \, \epsilon \frac{d\rho^{1halo}_{rad}}{dk'} V_{halo},
\ee
where $\frac{d\rho^{1halo}_{rad}}{dk'}$ is the spectral energy density of the excited radiation in the halo, which depends on the halo formation time\footnote{This is because the virialized energy density is roughly $\Delta_{vir}\rho_{bg}(z)$, which depends on the formation redshift and determines the value of the critical wavenumber $k_c$.} but is independent of the halo mass. $V_{halo}$ is the volume of the halo and $\epsilon$ is the escape fraction of the resonant radiation. 

Then, we can calculate the total resonant radiation energy density at $k_{21}$ by integrating over $t'$
\be
\frac{d\rho_{rad}}{dk}(k_{21}) =  \int_{t_i}^{t} dt' \frac{d\rho_{rad}}{dk'dt'} \bigg|_{k'=k_{21}\left(\tfrac{z'+1}{z+1}\right)} \bigg(\frac{z+1}{z'+1}\bigg)^3. \label{eq-drhodk}
\ee
Note that the factor $(\frac{z+1}{z'+1})^3$ accounts for the redshift of the spectral energy density due to the expansion of the universe.

For the excited radiation in halos, both energy cascade and thermalization may contribute to broadening the initially narrow spectrum. In what follows, I consider these two mechanisms separately.

\subsubsection{Energy cascade spectrum}
\label{sec3-2-1}

For the energy cascade spectrum, the injection energy is
\begin{align}
\frac{dE}{dk'} \, &= \, \epsilon \frac{f\rho_{axion}}{k_c}\bigg(\frac{k_c}{k'}\bigg)^n \frac{M}{\rho_{axion}} \nn
&= \, \epsilon f M k_c^{n-1} k'^{-n},
\end{align}
where 
\begin{align} 
k_c(t') \, &= \, 4\gphi m\phi_0(t')\\
&= 4 \gphi \big[ 2\Delta_{vir}\rho_{bg}(t_0)(z'+1)^3 \big]^{1/2},
\end{align}
which scales as $(z'+1)^{3/2}$ so we can rewrite the critical wavenumber as $k_c = k_{c,0}(z'+1)^{3/2}$ and $k_{c,0} = \gt\, 6\times10^{-24}eV$ is the critical wavenumber at $t_0$.
Note that the escape fraction $\epsilon$ and the radiation energy fraction $f$ are degenerate in the expression of $\frac{dE}{dk'}$. Thus, I redefine the escaped radiation energy fraction as
\be
\tilde{f} \equiv \epsilon f
\ee
and derive the constraint on this parameter in what follows.

The corresponding global radiation energy density is
\begin{widetext}
\begin{align}
\frac{d\rho_{rad}}{dk_{21}} \, &= \, \tilde{f} \int dt' \int dM M \frac{dn}{dMdt'} k_{c,0}^{n-1} k_{21}^{-n} (z'+1)^{(n-9)/2}(z+1)^{n+3}.
\end{align}
\end{widetext}

\begin{figure}[t]
  	\includegraphics[width=7.5cm]{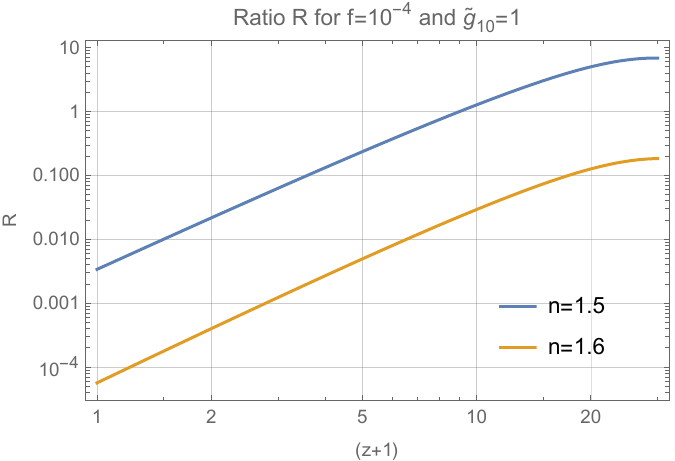}
	\caption{The ratio $\cR$ as a function of observed reshift $z$. In this figure, we set $\gt=1$ and $f=10^{-4}$.}
	\label{fig3}
\end{figure}

\begin{figure}[t]
  	\includegraphics[width=7cm]{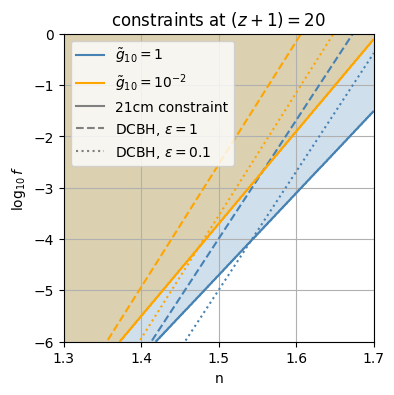}
	\caption{Constraints on the parametric resonance in halos from the global 21cm absorption feature at $(z+1)=20$. The blue and orange shaded regions represent the excluded parameter space for $\gt=1$ and $10^{-2}$, respectively. The dashed and dotted lines indicate the lower bounds on the radiation energy fraction required to generate sufficient LW radiation for the formation of DCBHs \cite{JH-DCBH}, assuming escape fraction $\epsilon=1$, and $0.1$, respectively.}
	\label{fig4}
\end{figure}

\begin{figure}[t]
  	\includegraphics[width=7cm]{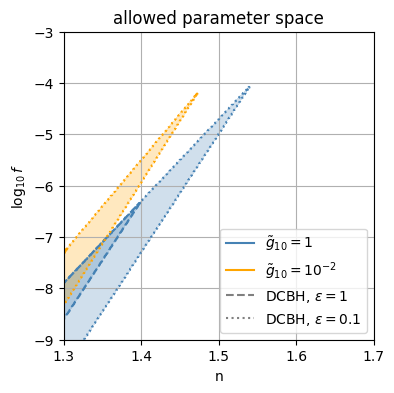}
	\caption{The shaded regions indicate the allowed parameter space where the formation of DCBHs in the scenario presented in \cite{JH-DCBH} is consistent with global 21cm constraints, under the assumption of an energy cascade spectrum. The color scheme and line styles are the same as in Fig.~\ref{fig4}. Note that the range of the vertical axis is 3 orders of magnitude smaller than that in the previous figures.}
	\label{fig5}
\end{figure}

This integral can be computed numerically, and the redshift dependence of the corresponding ratio $\cR$ is shown in Fig.~\ref{fig3}. Similar to the case of global resonant radiation, the impact of this scenario on 21cm signal is more pronounced at higher redshifts.  Therefore, I still adopt $(z+1)=20$ as the representative redshift for the constraints. 

The resulting constraints on $n$ and $\tilde{f}$ are presented in Fig.~\ref{fig4}, where shaded regions are excluded by observations. Note that the vertical axis is $\tilde f=\epsilon f$, which should be smaller than or equal to the fraction $f$ considered in the previous section.

The parametric resonance could play a role in the formation scenario of DCBHs \cite{JH-DCBH}, which requires strong LW radiation to suppress the abundance of $H_2$ in halos. To satisfy the DCBH formation criteria, there is a lower bound on the radiation fraction $\tilde{f}$, shown as the dashed and dotted lines for $\epsilon=1$ and $0.1$, respectively. In Fig.~\ref{fig4}, there exist narrow allowed regions in the parameter space, where the DCBH formation scenario does not conflict with the global 21cm observations. A smaller value of the escape fraction broadens the viable region. To better illustrate this, Fig.~\ref{fig5} shows the allowed parameter regions.

However, it should be noted that the escape fraction of radio photons is typically very high, i.e., $\epsilon\simeq1$, so it is unlikely for $\epsilon$ to be as low as $0.1$. Therefore, the DCBH formation scenario with resonant radiation requires a relatively small value of the index $n$ in order to remain consistent with the global 21cm observations.

\subsubsection{Thermalized spectrum}
\label{sec3-2-2}

In contrast to the global radiation case, thermalization of the excited radiation in virialized halos is reasonable and could also play a role in the DCBH formation process \cite{JH-DCBH}. In this subsection, I discuss the constraints for resonant radiation with a thermalized spectrum. In this case, 
\begin{align}
\frac{dE}{dk'} \,&=\, \frac{k'^3}{4\pi^3}\frac{1}{e^{k'/T_{rad}}-1} \cdot \frac{M}{\Delta_{vir}\rho_{bg,0}(z'+1)^3},
\end{align}
where $T_{rad}(f,z') = \big[ f \Delta_{vir}\rho_{bg,0}(z'+1)^3 \big]^{1/4}$ is the radiation temperature, which is significantly larger than $k_{21}'$ at the redshifts of interest. Therefore, we can approximate
\begin{align}
\frac{dE}{dk'} \,&\simeq\, \frac{k'^2}{4\pi^3} \frac{T_{rad} M}{\Delta_{vir}\rho_{bg,0}(z'+1)^3}.
\end{align}
Inserting this into eq.~\eqref{eq-drhodk}, we have
\begin{widetext}
\begin{align}
\frac{d\rho_{rad}}{dk_{21}} \, &= \, \frac{\epsilon f^{1/4}}{4\pi^3} \int dt' \int dM M \frac{dn}{dMdt'} k_{21}^2  \Delta_{vir}^{-3/4}\rho_{bg,0}^{-3/4}(z'+1)^{-13/4} (z+1)
\end{align}
\end{widetext}
The explicit $z$-dependence in this expression is the same as that in $\frac{d\rho_{CMB}}{dk_{21}}$, so the redshift dependence of the ratio $\cR$ is determined by the integration over $t'$, and therefore increases with the observed time $t$. At $t_0$, the ratio is
\be
\cR \,=\, 0.17 \, \epsilon \, f^{1/4},
\ee
which is always less than 1 for any value of $\epsilon,\, f<1$. Thus, the global 21cm signal does not constrain the resonant radiation with a thermalized spectrum.

Note that observations of the global 21cm signal are only valid before reionization, so strictly speaking, we should not derive constraints at $z=0$. However, as $\cR$ increases over time, the constraints at higher reshifts are correspondingly weaker, and this point does not affect our conclusion.

\section{Conclusions and discussion}
\label{sec4}






In this work, I investigate constraints on the excited radiation induced by an oscillating dark matter field using global 21cm observations. A parametric resonance of the electromagnetic gauge field occurs when it is coupled to an oscillating pseudoscalar axion field through a Chern-Simons interaction, leading to an exponential growth of the electromagnetic field amplitude with a pronounced peak around a critical wavenumber\footnote{Note that this is the wavenumber in physical coordinates.} 
\be
k_c \equiv 4\gphi m\phi_0
\ee
in the radiation spectrum. The initially narrow spectrum can then be broadened through an energy cascade or thermalization. The former mechanism results in a power-law energy spectrum, while the latter yields a blackbody spectrum.

Parametric resonance may occur globally soon after recombination or locally in virialized halos, providing potential explanations for the primordial magnetic field \cite{RB-magnetic} and the formation of SMBH seeds \cite{JH-DCBH}, respectively. In the previous section, I discussed the constraints on the two scenarios separately.

For the global resonance, we focus on the power-law spectrum, as thermalization is unlikely to occur on cosmological scales after recombination. The constraints on the radiation energy fraction $f$ from the global 21cm observations become more stringent with decreasing spectral index $n$ and increasing coupling constant $\gphi$, which is generally applicable to excited radiation with an energy-cascade spectrum.
The scenario for primordial magnetic field generation remains consistent with these bounds across a broad region of parameter space, as illustrated in Fig.~\ref{fig2}.

\begin{figure}[t]
  	\includegraphics[width=7cm]{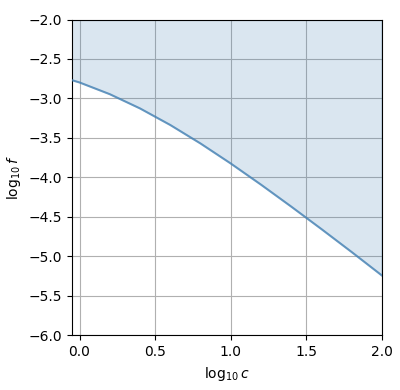}
	\caption{The shaded regions indicate the parameter space where the DCBH formation criteria in the scenario presented in \cite{JH-DCBH} are satisfied, assuming a thermalized spectrum. Note that the global 21cm signal does not constrain this mechanism, so this region is fully allowed.}
	\label{fig6}
\end{figure}

On the other hand, both energy cascade and thermalization are plausible in halos, and we examine the corresponding constraints for each mechanism. In the case of energy cascade, Fig. \ref{fig5} shows the parameter space allowed by both the global 21cm constraints and the DCBH formation criteria. However, this viable region is quite narrow and only exists for sufficiently small values of $n$.
In the case of thermalization, the 21cm signal does not impose constraints on the parameters.
For ease of comparison, Fig.~\ref{fig6} illustrates the region of the $c$-$f$ parameter space that permits DCBH formation, where $c$ denotes the halo concentration following the Navarro-Frenk-White density profile \cite{NFW}.

\begin{figure}[t]
  	\includegraphics[width=8.5cm]{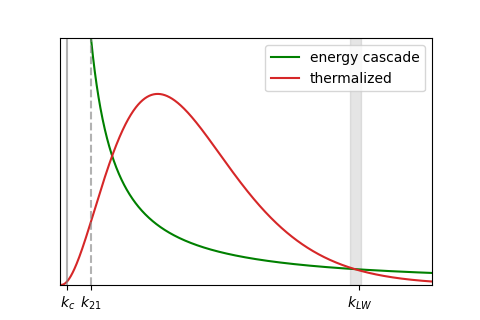}
	\caption{Sketch of the radiation spectrum with the energy cascade and thermalization, and the relative position of several typical photon energies in the parametric resonance scenario.}
	\label{fig7}
\end{figure}

From the previous analysis, we find that the constraints obtained under the energy cascade spectrum are consistently much tighter than those derived using the thermalized spectrum. This result is physically intuitive -- while the energy cascade spectrum monotonically decreases with the increase of $k$, the blackbody spectrum increases with $k$ in the low-energy regime, which encompasses the 21cm photon energy (see Fig.~ref{fig7}). Therefore, when the two spectra have comparable strength around the wavenumber of our interest in previous works, e.g. in the LW energy band, the thermalized spectrum leads to a significantly weaker signal at 21cm, resulting in looser observational constraints.

It should be noted that this work focuses on broad resonance, which arises when the oscillatory term $\pm4\gphi k \dot\phi A_k$ in the equation of motion \eqref{eq-A-EOM} dominates. This is in contrast to narrow resonance, which corresponds to the narrow instability bands of the Mathieu equation \cite{narrow-resonance1,narrow-resonance2}. In the case of broad resonance, most modes with $k<k_c$ are exponentially amplified, and the peak in the radiation spectrum arises from the $k$-dependence of the Floquet exponent and the fact that the initial phase space of modes scales as $k^3$. Thus, the dominant growing mode $k_c$ is a reasonable approximation in the majority of the broad resonance region. More detailed analyses of the precise fastest-growing mode and the impact of the inhomogeneous axion field distribution in halos are currently under investigation.

\section*{Acknowledgements}

\noindent  This research is supported by the Milton Leung Fellowship in Science and IBS under the project code IBS-R018-D3. I am deeply grateful to Robert Brandenberger and Vahid Kamali for many useful discussions and valuable suggestions that greatly contributed to the completion of this work.



\begin{thebibliography}{99}

\bibitem{axion-1}
R.~D.~Peccei, 
``The Strong CP Problem and Axions'', 
Axions: Theory, Cosmology, and Experimental Searches.
Berlin Springer Verlag Vol. 741, p. 3, 2008, [hep-ph/0607268].

\bibitem{axion-2}
J.~E.~Kim and G.~Carosi, 
``Axions and the strong  problem,''
Reviews of Modern Physics 82, 557 (2010).

\bibitem{axion-3}
D.~J.~E.~Marsh,
``Axion Cosmology,''
Phys. Rept. \textbf{643}, 1-79 (2016)
doi:10.1016/j.physrep.2016.06.005
[arXiv:1510.07633 [astro-ph.CO]].

\bibitem{axion-4}
P.~Sikivie, ``Axion Cosmology,'' 
Axions: Theory, Cosmology, and Experimental Searches.
 Berlin, Heidelberg: Springer Berlin Heidelberg, 2008. 19-50.


\bibitem{FDM-1}
L~Hui, J.~P.~Ostriker, S.~Tremaine, E.~Witten,
``Wave Dark Matter,''
Ann.Rev.Astron.Astrophys. \tbf{59} (2021) 247-289.

\bibitem{FDM-2}
L~Hui, J.~P.~Ostriker, S.~Tremaine, E.~Witten,
``Ultralight scalars as cosmological dark matter,''
Phys.Rev.D \tbf{95} (2017) 4, 043541.

\bibitem{FDM-3}
E.~Ferreira,
``Ultra-light dark matter,''
Astron.Astrophys.Rev. \tbf{29} (2021) 1, 7.


\bibitem{RB-misalignment}
R.~H.~Brandenberger,
``A Non-Inflationary Axion and ALP Misalignment Mechanism,''
arXiv:2507.00785 [hep-ph].

\bibitem{misalignment-2}
M. Dine, 
``The Problem of Axion Quality: A Low Energy Effective Action Perspective,'' 
[arXiv:2207.01068 [hepph]].

\bibitem{misalignment-3}
J.~Preskill, M.~B.~Wise, and F.~ Wilczek,
``Cosmology of the Invisible Axion'',
Published in Phys.Lett.B 120 (1983) 127.

\bibitem{misalignment-4}
L.F. Abbott, and P. Sikivie,
``A Cosmological Bound on the Invisible Axion'',
Phys.Lett.B 120 (1983) 133.

\bibitem{misalignment-5}
M. Dine, and W. Fischler, 
``The Not So Harmless Axion'',
Phys.Lett.B 120 (1983) 137.

\bibitem{violate-1}
S.~M.~Carroll, G.~B.~Field and R.~Jackiw, 
``Limits on a Lorentz and Parity Violating Modification of Electrodynamics,'' 
Phys. Rev. D \tbf{41}, 1231 (1990) doi:10.1103/PhysRevD.41.1231

\bibitem{violate-2}
W.~D.~Garretson, G.~B.~Field, and S.~M.~Carroll, 
``Primordial magnetic fields from pseudoGoldstone bosons,'' 
Phys. Rev. D \tbf{46}, 5346-5351 (1992) doi:10.1103/PhysRevD.46.5346 [arXiv:hep-ph/9209238 [hep-ph]].

\bibitem{violate-3}
G.~B.~Field and S.~M.~Carroll, 
``Cosmological magnetic fields from primordial helicity,'' 
Phys. Rev. D \tbf{62}, 103008 (2000) doi:10.1103/PhysRevD.62.103008 [arXiv:astro-ph/9811206 [astro-ph]].

\bibitem{JF1}
J.~Frohlich and B.~Pedrini,
``New applications of the chiral anomaly,''
[arXiv:hep-th/0002195 [hep-th]].

\bibitem{JF2}
J.~Frohlich and B.~Pedrini,
``Axions, quantum mechanical pumping, and primeval magnetic fields,''
[arXiv:cond-mat/0201236 [cond-mat]].

\bibitem{Joyce}
M.~Joyce and M.~E.~Shaposhnikov,
``Primordial magnetic fields, right-handed electrons, and the Abelian anomaly,''
Phys. Rev. Lett. \textbf{79}, 1193-1196 (1997)
doi:10.1103/PhysRevLett.79.1193
[arXiv:astro-ph/9703005 [astro-ph]].

\bibitem{RB-magnetic}
R.~Brandenberger, J.~Fr\"ohlich, H.~Jiao,
``Cosmological Magnetic Fields from Ultralight Dark Matter,''
[arXiv:2502.19310 [hep-ph]].



\bibitem{Evan1}
E.~McDonough, H.~Bazrafshan Moghaddam and R.~H.~Brandenberger,
``Preheating and Entropy Perturbations in Axion Monodromy Inflation,''
JCAP \textbf{05}, 012 (2016)
doi:10.1088/1475-7516/2016/05/012
[arXiv:1601.07749 [hep-th]].

\bibitem{Evan2}
H.~Bazrafshan Moghaddam, E.~McDonough, R.~Namba and R.~H.~Brandenberger,
``Inflationary magneto-(non)genesis, increasing kinetic couplings, and the strong coupling problem,''
Class. Quant. Grav. \textbf{35}, no.10, 105015 (2018)
doi:10.1088/1361-6382/aaba22
[arXiv:1707.05820 [astro-ph.CO]].

\bibitem{Rudnei}
R.~Brandenberger, V.~Kamali and R.~O.~Ramos,
``Decay of ALP condensates via gravitation-induced resonance,''
JCAP \textbf{11}, 009 (2023)
doi:10.1088/1475-7516/2023/11/009
[arXiv:2303.14800 [hep-ph]].


\bibitem{JH-DCBH}
H.~Jiao, R.~Brandenberger, and V.~Kamali,
``Direct Collapse Supermassive Black Holes from Ultralight Dark Matter,''
[arXiv:2503.19414 [astro-ph.CO]].


\bibitem{PMF-1}
T.~Vachaspati, 
``Progress on cosmological magnetic fields,'' 
Rept. Prog. Phys. \tbf{84}, no.7, 074901 (2021) doi:10.1088/1361-6633/ac03a9 [arXiv:2010.10525 [astroph.CO]].

\bibitem{PMF-2}
R.~Durrer and A.~Neronov, 
``Cosmological Magnetic Fields: Their Generation, Evolution and Observation,'' 
Astron. Astrophys. Rev. \tbf{21}, 62 (2013) doi:10.1007/s00159-013-0062-7 [arXiv:1303.7121 [astroph.CO]].

\bibitem{PMF-gen-1}
M.~M.~Anber and L.~Sorbo, 
``N-flationary magnetic fields,'' 
JCAP \tbf{10}, 018 (2006) doi:10.1088/14757516/2006/10/018 [arXiv:astro-ph/0606534 [astro-ph]].

\bibitem{PMF-gen-2}
R. Durrer, L. Hollenstein and R. K. Jain, 
``Can slow roll inflation induce relevant helical magnetic fields?,'' 
JCAP \tbf{03}, 037 (2011) doi:10.1088/1475-7516/2011/03/037 [arXiv:1005.5322 [astro-ph.CO]].

\bibitem{PMF-gen-3}
T. Fujita, R. Namba, Y. Tada, N. Takeda and H. Tashiro, 
``Consistent generation of magnetic fields in axion inflation models,'' 
JCAP \tbf{05}, 054 (2015) doi:10.1088/14757516/2015/05/054 [arXiv:1503.05802 [astro-ph.CO]].

\bibitem{PMF-gen-4}
L.~Campanelli, 
``Helical Magnetic Fields from Inflation,'' 
Int. J. Mod. Phys. D \tbf{18}, 1395-1411 (2009) doi:10.1142/S0218271809015175 [arXiv:0805.0575 [astroph]].



\bibitem{AGN1}
X. Fan, M. A. Strauss, D. P. Schneider, R. H. Becker, R. L. White, {\it et al.},
``A survey of $z>5.7$ quasars in the Sloan Digital Sky Survey. II. Discovery of three additional quasars at $z>6$,
the Astronomical Journal, {\bf 125} (2003):1649.

\bibitem{AGN2}
C. J. Willott, {\it et al.}, 
``Four quasars above redshift 6 discovered by the Canada-France High-z Quasar Survey,'' 
The Astronomical Journal {\bf 134}.6 (2007): 2435.

\bibitem{AGN3}
J. Yang, F. Wang, A. J. Barth, {\it et al.},
``Probing Early Supermassive Black Hole Growth and Quasar Evolution with Near-infrared Spectroscopy of 37 Reionization-era Quasars at $6.3<z\leq7.64$,''
The Astrophysical Journal {\bf 923}.2 (2021): 262.

\bibitem{AGN4}
L. Jiang, N Kashikawa, S Wang, {\it et al.},
``Evidence for GN-z11 as a luminous galaxy at redshift 10.957,'' 
Nature Astronomy {\bf 5}.3 (2021): 256-261.

\bibitem{AGN5}
P. A. Oesch, {\it et al.}, 
``A remarkably luminous galaxy at $z= 11.1$ measured with Hubble space telescope Grism spectroscopy,''
The Astrophysical Journal {\bf 819}.2 (2016): 129.


\bibitem{DCBH-1}
V. Bromm and A. Loeb, 
``Formation of the first supermassive black holes'' 
Astrophys.J. \tbf{596} (2003) 34-46, [arXiv:astro-ph/0212400 [astro-ph]]

\bibitem{DCBH-2}
M. Umemura, A. Loeb, and E.L. Turner, 
``Early cosmic formation of massive black holes'', 
Astrophys.J. \tbf{419} (1993) 459, [arXiv:astro-ph/9303004 [astro-ph]]

\bibitem{DCBH-3}
A. Loeb and F.A. Rasio, 
``Collapse of primordial gas clouds and the formation of quasar black holes'', 
Astrophys.J. \tbf{432} (1994) 52, [arXiv:astro-ph/9401026 [astroph]]

\bibitem{DCBH-4}
D.J. Eisenstein and A. Loeb, 
``Origin of quasar progenitors from the collapse of low spin cosmological perturbations'', 
Astrophys.J. \tbf{443} (1995) 11, [arXiv:astroph/9401016 [astro-ph]]

\bibitem{DCBH-5}
K. Inayoshi, E. Visbal, and Z. Haiman, 
``The Assembly of the First Massive Black Holes'', 
Ann.Rev.Astron.Astrophys. \tbf{58} [arXiv:1911.05791 [astro-ph.GA]]

\bibitem{DCBH-6}
S.P. Oh and Z. Haiman, 
``Second-generation objects in the universe: radiative cooling and collapse of halos with virial temperatures above $10^4$ kelvin'', 
Astrophys.J. \tbf{569} (2002) 558, [arXiv:astro-ph/0108071 [astro-ph]]

\bibitem{DCBH-7}
Z. Haiman, T. Abel, and M.J. Rees, 
``The radiative feedback of the first cosmological objects'', 
Astrophys.J. \tbf{534} (2000) 11-24, [arXiv:astro-ph/9903336 [astro-ph]]

\bibitem{DCBH-8}
Agarwal, Bhaskar, et al. 
``New constraints on direct collapse black hole formation in the early Universe'', 
Mon.Not.Roy.Astron.Soc. \tbf{459} (2016) 4209-4217.

\bibitem{DCBH-9}
Y. Lu, Z. S. C. Picker, and A. Kusenko, 
``Direct collapse supermassive black holes from relic particle decay'', 
Physical Review Letters \tbf{133} (2024) 091001.


\bibitem{21cm-1}
J.~Pritchard, and A.~Loeb, 
``21-cm cosmology,''
Rept.Prog.Phys. \tbf{75} (2012) 086901.

\bibitem{21cm-2}
C.~Feng, G.~Holder
``Enhanced global signal of neutral hydrogen due to excess radiation at cosmic dawn,''
Astrophys.J.Lett. \tbf{858} (2018) 2, L17

\bibitem{EDGES}
J.~D.~Bowman, A.~E.~E.~Rogers, R.~A.~Monsalve, T.~J.~Mozdzen and N.~Mahesh, 
``An absorption profile centred at 78 megahertz in the sky-averaged spectrum,'' 
Nature \tbf{555}, no. 7694, 67 (2018) [arXiv:1810.05912 [astro-ph.CO]].

\bibitem{EDGES-problem}
S.~Singh, J.~Nambissan T., R.~Subrahmanyan, and et al., 
``On the detection of a cosmic dawn signal in the radio background,'' 
Nat Astron \tbf{6}, 607-617 (2022).



\bibitem{axionphoton-1}
Ciaran O'Hare, ``cajohare/AxionLimits: AxionLimits,'' http://cajihare.github.io/AxionLimits/.


\bibitem{AxionMass1}
K.~K.~Rogers and H.~V.~Peiris,
``Strong Bound on Canonical Ultralight Axion Dark Matter from the Lyman-Alpha Forest,''
Phys. Rev. Lett. \textbf{126}, no.7, 071302 (2021)
doi:10.1103/PhysRevLett.126.071302
[arXiv:2007.12705 [astro-ph.CO]]. 

\bibitem{AxionMass2}
D.~Y.~Cheong, N.~L.~Rodd and L.~T.~Wang,
``A Quantum Description of Wave Dark Matter,''
[arXiv:2408.04696 [hep-ph]].


\bibitem{magnetic-1}
R.~Durrer and A.~Neronov, 
``Cosmological Magnetic Fields: Their Generation, Evolution and Observation,'' 
Astron. Astrophys. Rev. \tbf{21}, 62 (2013) 
doi:10.1007/s00159-013-0062-7 
[arXiv:1303.7121 [astroph.CO]].


\bibitem{Kolmogorov}
A. N. Kolmogorov,
``A refinement of previous hypotheses concerning the local structure of turbulence in a viscous incompressible fluid at high Reynolds number''. 
Journal of Fluid Mechanics. 13 (1), 82 (1961).

\bibitem{Batchelor}
G.K.  Batchelor, 
``Small-scale variation of convected quantities like temperature in turbulent fluid. Part 1. General discussion and the case of small conductivity'', 
Journal of Fluid Mechanics, 5,  113,  (1959).

\bibitem{NFW}
J.~F.~Navarro, C.~S.~Frenk and S.~D.~M.~White,
``A Universal density profile from hierarchical clustering,''
Astrophys. J. \textbf{490}, 493-508 (1997)
doi:10.1086/304888
[arXiv:astro-ph/9611107 [astro-ph]].

\bibitem{narrow-resonance1}
R.~Brandenberger, P.~C.~M.~Delgado, A.~Ganz and C.~Lin,
``Graviton to photon conversion via parametric resonance,''
Phys. Dark Univ. \textbf{40}, 101202 (2023)
doi:10.1016/j.dark.2023.101202
[arXiv:2205.08767 [gr-qc]].

\bibitem{narrow-resonance2}
M.~P.~Hertzberg, and E.~D.~Schiappacasse, 
``Dark matter axion clump resonance of photons,'' 
JCAP \tbf{11}, 004 (2018).


\end{thebibliography}
\end{document}